# Vapor-like liquid densities affect the extension of the critical point's influence zone

Jose Luis Rivera,[*] Homero Nicanor-Guzman, and Roberto Guerra-Gonzalez

*Facultad de Ingeniería Química, Universidad Michoacana de San Nicolás de Hidalgo, Morelia, Michoacán 58000, México*

Abstract

*The critical point affects the coexistence behavior of the vapor-liquid equilibrium densities. The length of the critical influence zone is under debate because for some properties, like shear viscosity, the extension is only a few degrees, while for others, such as the density order parameter, the critical influence zone range covers up to hundreds of degrees below the critical temperature. Here we show that for a simple molecular potential of ethane, the critical influence zone covers a wide zone of tens of degrees (below the critical temperature) down to a transition temperature, at which the apparent critical influence zone vanishes and the transition temperature can be predicted through a pressure analysis of the coexisting bulk liquid phase. The liquid phases within the apparent critical influence zone show low densities, making them behave internally like their corresponding vapor phases. Therefore, the experimentally observed wide extension*

---

[*] Corresponding Author. Email: jlrivera@umich.mx

*of the critical influence zone is due to a vapor-like effect due to low bulk liquid phase densities.*

1. INTRODUCTION

Vapor – liquid equilibrium (VLE) for pure components is present between the triple and critical points; at a transition temperature between these two points, and up to the critical point (critical influence zone), equilibrium liquid phases start to behave more like their corresponding vapor phases,[1] characterized by larger density changes as the temperature increases. In the critical influence zone (CIZ), the coexisting vapor and liquid densities approach asymptotically, indicating a strong influence of the critical behavior on the coexisting densities, and the universal power law describes their difference:

$$\Delta\rho = B_0 \tau^\beta \qquad (1)$$

where $\Delta\rho = \rho_l - \rho_v$ is the order parameter of phase transition, $\tau = (T_C - T)/T_C$, $T_C$ is the critical temperature, $\beta$ is a universal critical exponent, and $B_0$ is a constant dependent on the system. After a threshold temperature, $\Delta\rho$ deviates from the universal power law, and the correct value of $\Delta\rho$ can be calculated by introducing non-asymptotic corrections.[2,3] Water order-parameters (experimental) show a linear behavior and follow Eq. 1 down to ~130° below $T_C$. In comparison, a pure Lennard-Jones fluid follows the universal power law for almost all its coexistence region.[4] The wide influence zone in the order parameter present in water is unexpected, since the critical influence on other properties (shear viscosity) only covers a few degrees below $T_C$.[4,5] Clearly, the main difference between the Lennard-Jones fluid and water is the long intermolecular interactions present in the water molecule, but a second difference between these two systems is the presence of intramolecular interactions in the water molecule. Intra- and inter-molecular forces are not completely independent in equilibrium phases; for a specific fluid, its liquid phases in equilibrium close to the triple point will have in average more compressed molecules (higher intramolecular forces), than liquid phases in equilibrium near the critical point, which will have lower densities, and therefore the molecules are less compressed. In the other hand, molecules in liquid phases close to the triple point will show stronger intermolecular forces, as those systems are denser than systems at higher temperatures. Therefore, intra- and inter-molecular forces are correlated as we move from the triple to the critical point.

To understand the unexpected range of the critical-influence in the order parameter of water, we analyzed through Molecular Dynamics (MD) simulations the ethane molecule, which compared to a Lennard-Jones fluid shows intramolecular interactions as the water molecule, but it has less complex intermolecular interactions than the water molecule. For the ethane molecule, experimental values of $\Delta\rho$ as a function of $\tau$ have been plotted in Figure 1. The critical behavior apparently influences $\Delta\rho$ for ~75º, between a transition temperature ~230K and the $T_C$. For this range of temperatures, $\Delta\rho$ values show good agreement with the universal power law (Eq. 1).

In this paper we study the intramolecular forces in the pressure profiles of vapor – liquid systems of ethane under equilibrium for the whole range of coexistence, and correlate the changes in the bulk liquid pressure profiles to their corresponding vapor phases in the critical influence zone to delimitate its range of influence. The remainder of the article is organized as follows: in section II, we provide details of the molecular models, and the molecular simulation methodologies used to produce the results discussed in section III, and finally we report the conclusions in section IV.

II. METHODOLOGY

MD simulations allows the study of systems under phase equilibrium, and specifically can be applied to examine the contributions of the intramolecular and intermolecular interactions to the thermophysical properties directly at the equilibrium interface, and their corresponding bulk phases.[6-10] The simulation of the coexisting VLE uses pair-effective intermolecular potentials that are capable of reproducing their thermophysical properties,[11-13] and transport properties for simple fluids.[14-16] MD simulations in VLE for ethane show high- and low-density coexisting phases divided by interfaces (Figure 2, top). Density profiles can be obtained from time averages on mechanically stable configurations to define the bulk and interfacial zones (Figure 2, bottom). The bulk coexisting densities from the simulations results using the MIE potential used in this work[11] reproduce well the experimental coexistence behavior[17] (Figure 3). Lennard-Jones based models reproduce as well the coexisting curve.[18, 19]

The VLE of ethane was simulated using the MD methodology (in-house code) in the range of 100 - 260 K. The simulations were carried out using the Verlet algorithm with a Nose thermostat,[20] and a time step of 1 fs. A cutoff radius of 4.5 sigmas (17.0235 Å) was employed, which is long enough to avoid the use of long-range corrections.[21] Some points were recalculated using longer cutoff radius (5.5 sigmas) to verify the employed radius. The initial system consists of a cubic cell of ethane molecules surrounded by three empty rectangular cells (Figure 1). The dimensions of the combined simulation cell are 43.7 x 43.7 x 174.8 Å$^3$, and the cell contains 1000 molecules of ethane. We use 10 ns to generate statistical averages of the properties, following an initial run of 1 ns for equilibration. To equilibrate the system, it was slowly heated, starting from a configuration at 0 K, in order to avoid translation of the "liquid slab" of ethane through the simulation cell. To avoid oscillations problems in the calculated properties, equilibration was carried out using the Canonical ensemble, while production was carried out at the microcanonical ensemble.[22] Density and pressure profiles were calculated using the standard expressions reported elsewhere.[10, 23] The interaction potential for ethane molecules was the flexible version of the MIE potential,[11] which reproduces well the coexistence properties, including the coexisting densities, critical point, and vaporization enthalpy.[11] The original MIE potential for ethane is rigid, but the inclusion of flexible bonds does not affect their ability to reproduce the coexistence properties.[11] The vibrational bond form used in this work is a quadratic harmonic oscillator, similar to the one used in the flexible TraPPE potential.[13]

III. RESULTS

The influence of the intramolecular interactions on the length of the critical-like behavior can be studied through an analysis of the bulk pressures in the VLE for a simple molecular potential of ethane. The viral expression of the pressure tensor can be divided into three contributions:

$$P_\alpha = (kinetic)_\alpha + \binom{intermolecular}{potential}_\alpha + \binom{intramolecular}{potential}_\alpha \quad (2)$$

where $P_\alpha$ represents the normal ($P_N$) or tangential pressure ($P_T$). Figure 4 show simulation results of the position profiles of the contributions to $P_N$, and $P_T$ at 100 K, respectively. The concept of pressure is commonly associated with intermolecular forces, but the inclusion of intramolecular contributions to the total pressure, accomplishes the requirement of mechanical equilibrium, making the total pressure profiles in the bulk vapor and bulk liquid phases equal. A previous report has shown the need of inclusion of intramolecular contributions to the pressure profile to make the profiles mechanically stable in rigid diatomic molecules.[24] In this work, we report for the first time the effects of intramolecular contributions of flexible molecules on the pressure profile. Kinetic pressure profiles show the same behavior in the normal and tangential profiles (Figures 5a and 5b). Intermolecular pressure profiles are shown in Figures 5e and 5f for $P_N$ and $P_T$, respectively. In the coexisting bulk liquid phase (CBLP), negative intermolecular contributions are the result of net cohesive forces, and the average pressure decreases as the temperature increases. In the coexisting bulk vapor, the intermolecular contributions are insignificant because of the long separations of the molecules in the vapor phase. At the interface, the pronounced peaks at low temperatures (100 and 140 K) in the tangential contributions are the main contributors to the surface tension. Surface tension results (Figure 6) using the MIE potential show good agreement with experimental results[17] at moderate and high temperatures, while at the low temperature of 100 K, there is a small deviation, but lower than other simple ethane models like OPLS and TraPPE.[18, 19]

Figures 5c and 5d show the intramolecular, normal and tangential pressure profiles, respectively; at low temperatures, the average intramolecular contribution in the CBLP is positive indicating that the majority of molecular states show compressed bond separations, which can be expected for a normal liquid state, but as the temperature increases, the average magnitude of the intramolecular contribution decreases, taking negative average values at higher temperatures. Negative intramolecular pressures are also found in the average magnitude of the coexisting bulk vapor phase, and they probably indicate that the majority of molecular states show bond separations with positive deviations from the ground value, which are probably the result of intramolecular pulling forces acting on the bond atoms, which are the result of net attractive intermolecular interactions, typical of bulk vapor phases.

The intramolecular pressure in the CBLP can be used to establish the transition temperature, at which the CBLPs move from positive to negative intramolecular pressures (similar to those present in a coexisting bulk vapor phase, Figure 5). Figure 7 shows the average intramolecular pressures in the CBLP as a function of the system temperature. The intramolecular pressure regularly reduces as the temperature increases, almost in a linear fashion. A quadratic regression shows the locations of the transition temperature ~224 K. This temperature is also represented in Figure 1, and falls within the transition range of temperatures where the CIZ vanishes. In Figure 7 we also observe that the average intermolecular pressure decreases monotonically, and together with the intramolecular pressure, the total potential pressure shows a minimum between 240 and 250 K. Compared to the potential pressure, the kinetic pressure shows a symmetric behavior, with a maximum also in the range of 240 – 250 K. The maximum kinetic pressure in the CBLP can be understood if we take into account that the kinetic pressure is a function not only of the increasing temperature, but also a function of the bulk liquid density, which decreases as the temperature increases, and the decrements are more pronounced at temperatures close to the critical point. Therefore, there should be a maximum in the kinetic pressure when increasing contributions due to temperature increments cannot compete with decreasing contributions due to increasingly lower densities. The reported values for the kinetic, intramolecular, and intermolecular pressures for ethane added into the total liquid pressure reproduces well the experimental values of the vapor pressure (Figure 9), which for a system under equilibrium should be the same.

The threshold value for the transition temperature where the CIZ vanishes can also be studied more directly using the bond-separation distributions as a function of the temperature. The insight of Figure 8 shows the bond-separation deviation (from the ground value) distributions at a low (100 K) and a high temperature (260 K), for the CBLP of ethane in the VLE. As expected, the distributions widen as the temperature increases, but small changes in the median value are apparent. Figure 8 shows the deviation of the median of the bond-separation distributions from the ground value ($b_0 = 1.54$ Å) as a function of the system temperature. The deviations grow from negative values in the range of 100 – 240 K to positive values in the range 240 - 260 K, with a value of zero at 240 K. The transition temperature (240 K) from negative to positive deviations also falls within the transition region plotted in Figure 1. Negative median deviations are expected in moderate and highly dense phases like compressed liquids and solids, where strong, cohesive forces compress the molecular volume, while positive median deviations are expected in low dense phases (vapors) where atoms in the molecular bond are pulled apart due to net attractive intermolecular forces. The net attractive intermolecular forces in low dense phases are the result of long intermolecular separations, typically found in vapors and gasses. A simple calculation also confirms the bond deviations behavior; at the transition temperature (240 K), the CBLP density produces a site volume (we have two $CH_3$ sites per ethane molecule) equal to the characteristic separation sigma (MIE potential), where the intermolecular interaction energy is zero, then for this delimitating site volume, molecules are packed without any compression in their bond separation due to intermolecular interactions.

We also investigated the effect of the cutoff radius, recalculating points in the threshold temperature where the CIZ vanishes using longer cutoff radius (5.5 sigmas) and found negligible effects on the determination of the transition temperature. We were also interested in the effects of the flexibility of the intramolecular potential used, and recalculated some points near the threshold temperature where the CIZ vanishes using an intramolecular potential 10 times stiffer, more similar to the behavior of the original MIE potential (rigid) and found that the threshold temperature move 2 degrees (~ 222 K).

IV. CONCLUSIONS

By using Molecular Dynamics simulations, can be predicted the transition temperature, which delimits the critical influence zone. We demonstrate that, the intramolecular pressure in the coexisting bulk liquid phase can be used to predict the extension of the apparent critical influence region. The coexisting bulk liquid phases in the apparent critical influence zone behaves more like vapors in terms of its average intramolecular forces. Within the temperatures studied, we found that coexisting bulk liquid phases below the critical influence zone show negative intermolecular pressures and positive intramolecular pressures, while for coexisting bulk liquid phases within the critical influence zone both pressures are negative. For vapor phases, the intramolecular part is negative, and the intermolecular part is also negative and almost zero. Bond distributions confirm the "vapor-like" behavior of coexisting bulk liquid phases in the critical influence zone, which for the ethane molecules extend in the range of ~ 230 K to the critical point. Liquids behaving like vapors have also been observed experimentally in calorimetric studies of drops with low densities.[25] Therefore, the apparent wide zone of the experimentally observed critical influence zone is due to two contributions; from the transition temperature to the critical point we have the vapor-like effect due to low coexisting bulk liquid phase densities, while at temperatures near the vicinity of the critical temperature, the critical effects dominate.

**Acknowledgments**

JLR thanks CONACYT – Ciencia Básica (México) under Grant No. 134508, and Universidad Michoacana de San Nicolás de Hidalgo for sabbatical support. HNG thanks UMSNH for a scholarship.

**Close Captions.**

**Figure 1.** Plot of the order parameter $\Delta\rho=\rho_l-\rho_v$ vs. the reduced temperature $\tau$ for real ethane[17] (solid circles). Results from the scaling equation (1) are shown by the straight solid line. Vertical dashed lines represent the transition temperatures from saturated liquids to vapor-like saturated liquids using the criteria of the intramolecular pressure contributions (left) and the criteria of the average value of the median bond separations (right). The dashed circle represents the region where the CIZ vanishes.

**Figure 2.** Density profiles of ethane in the VLE as a function of the simulation cell position in the inhomogeneous direction of the simulation cell (bottom). The black line represents the profile at 100 K, while blue and red represent the profiles at 180 K, and 260 K, respectively. The top figure represents an instantaneous snapshot of the ethane molecules in the vapor-liquid equilibrium at 260 K.

**Figure 3.** Coexistence curve of ethane in the vapor liquid-equilibrium. The continuous line represents the experimental results.[17] Plus symbols represent Monte Carlo Simulations results for the rigid MIE potential.[11] Circles represent Molecular Dynamics simulations results using the flexible MIE potential used in this work.

**Figure 4.** Profiles of the contributions to the normal (left) and tangential pressures (right), for ethane in the VLE as a function of the simulation cell position in the inhomogeneous direction of the simulation cell. Kinetic, intramolecular and intermolecular contributions for ethane at 100 K, are represented in the Figures by blue, red and green profiles, respectively. Black profiles represent the total values.

**Figure 5.** Profiles of the contributions to the normal (left) and tangential pressures (right), for ethane in the VLE as a function of the simulation cell position in the inhomogeneous direction of the simulation cell. The color code in Figures 5a – 5f represent results at 100 K (black), 140 K (green), 220 K (blue), and 260 K (red). Kinetic contributions are shown in Figures 5a and 5b, while intramolecular and intermolecular contributions are shown in Figures 5c – 5d, and 5e – 5f, respectively.

**Figure 6.** Surface tension of ethane as a function of the temperature. The dashed line represents experimental results.[17] The continuous line is a fitted result to the experimental results at low temperatures. Triangle, rectangles, and plus symbols represent simulations results using the OPLS, flexible TraPPE, and rigid TraPPE potentials, respectively.[18, 19] Circles represent results of this work using the flexible MIE potential.

**Figure 7.** Average values of the intramolecular (squares), intermolecular (circles), potential (sum of the intramolecular and intermolecular pressure, crosses), and kinetic (stars) pressures in the CBLP for ethane, as a function of the system temperature.

**Figure 8.** Median of the deviations of the bond separation distributions, for ethane in the VLE as a function of the system temperature. Bond separation distributions for ethane in the VLE, at 100 K (red circles), and 260 K (blue circles) (insight of Figure 8).

**Figure 9.** Vapor pressures as a function of the system temperature. Dotted line represent experimental results.[17] Circles represent results of this work using the MIE potential.[11]

**Figure 1**

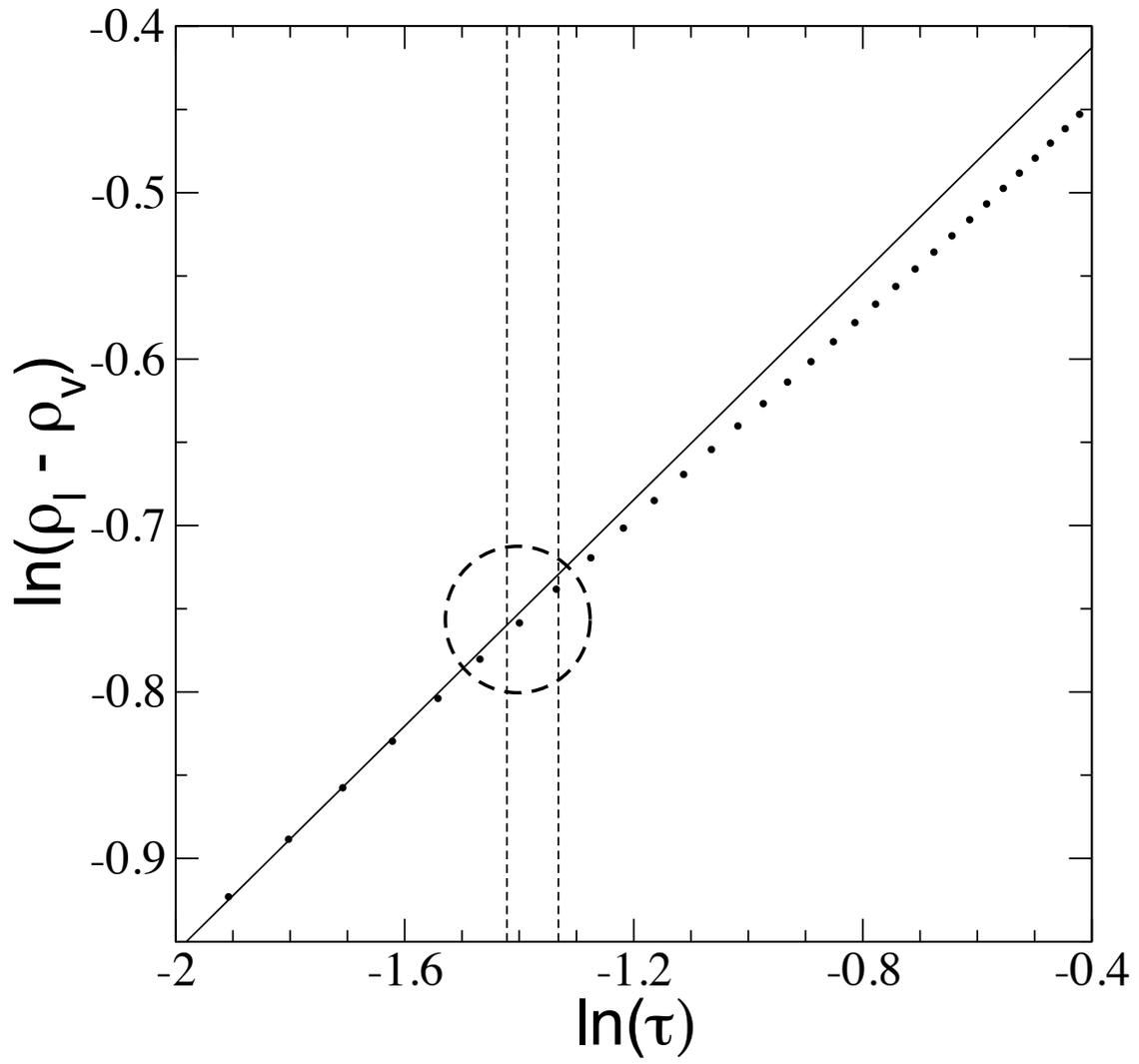

**Figure 2**

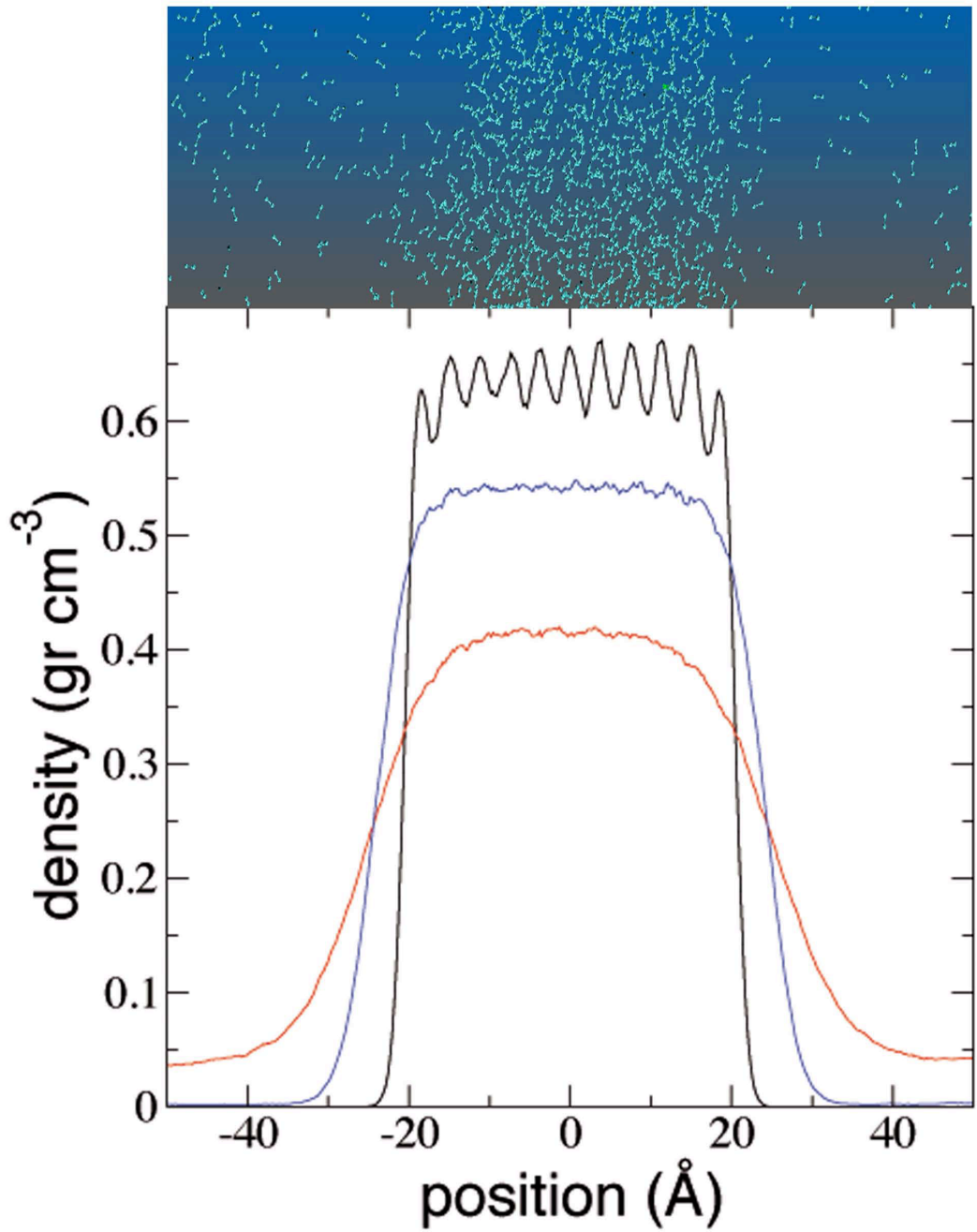

**Figure 3**

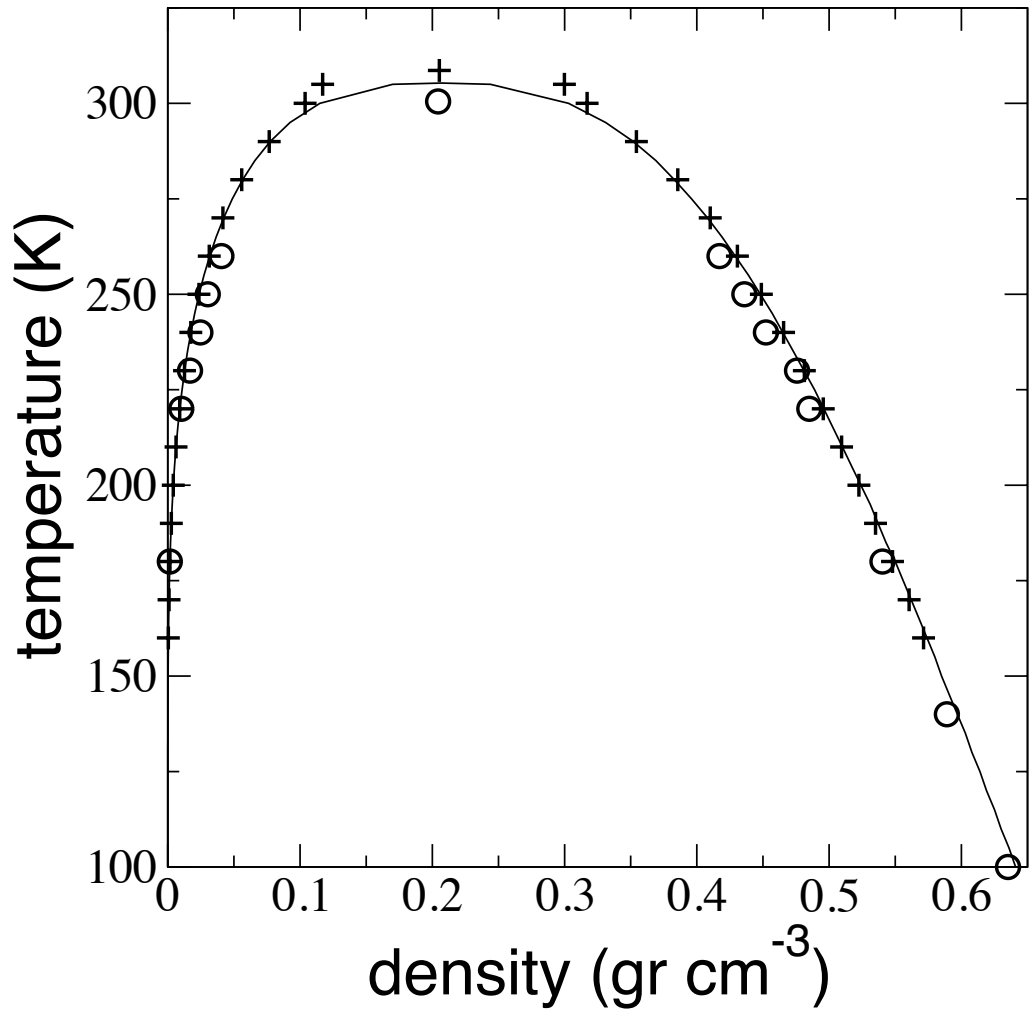

**Figure 4**

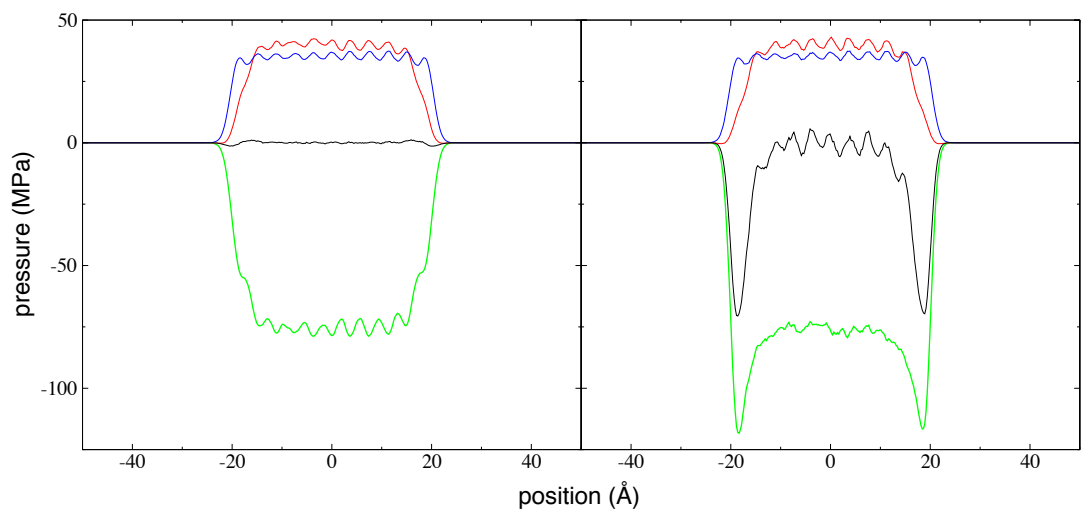

**Figure 5**

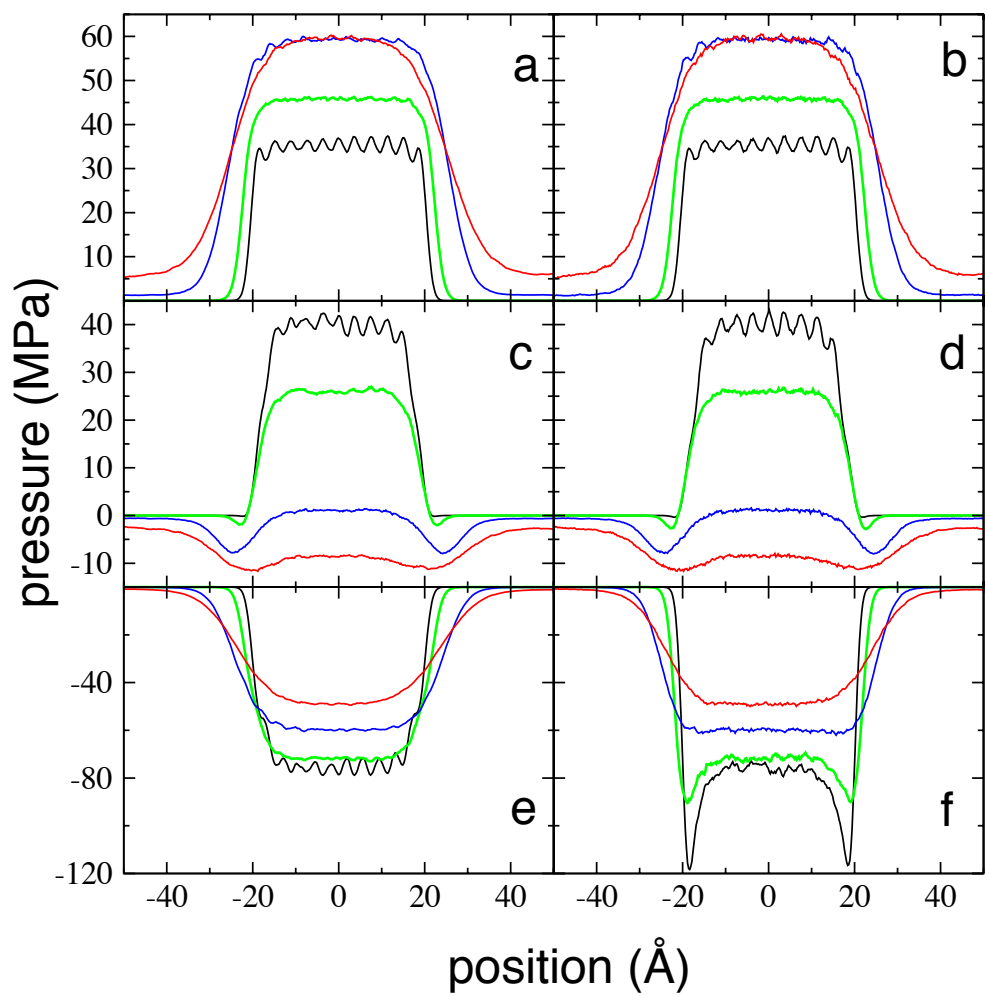

**Figure 6**

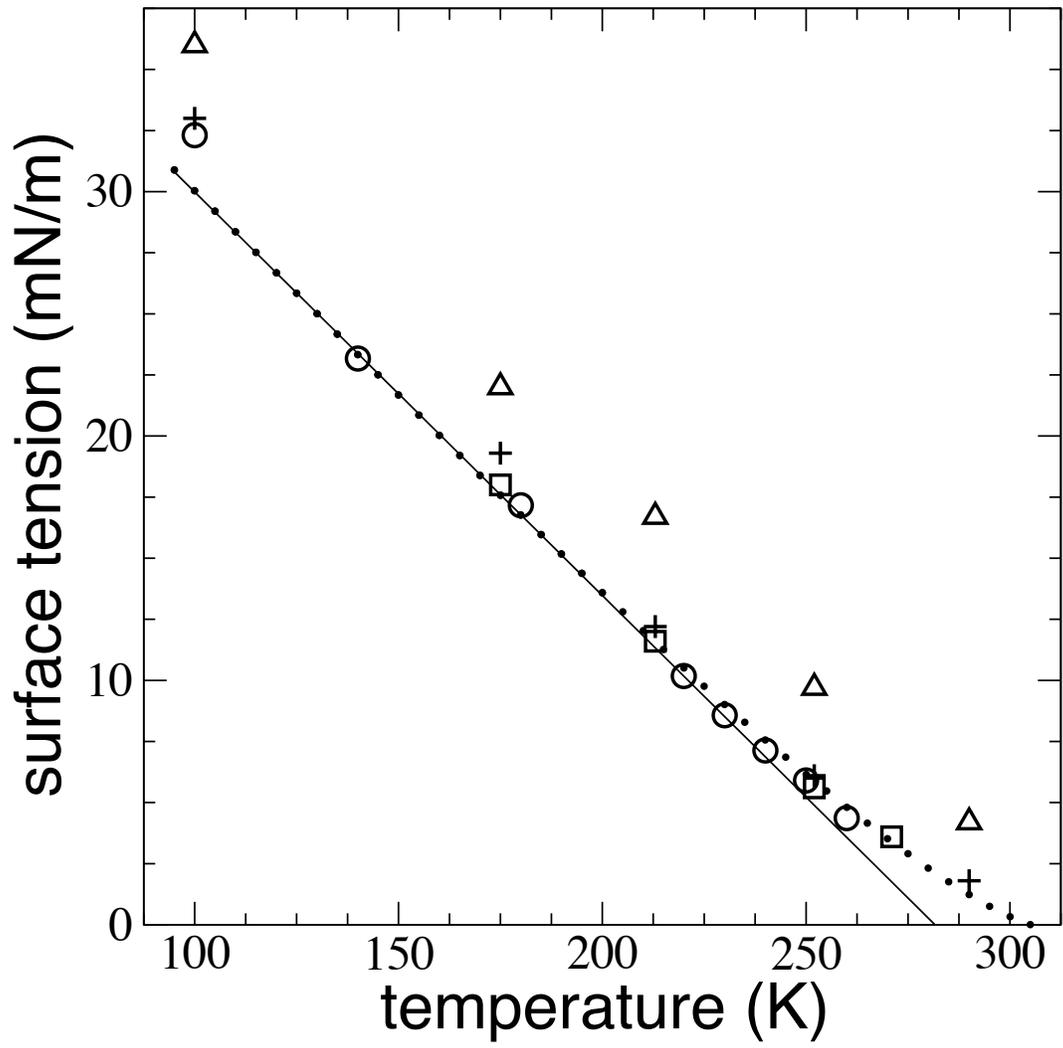

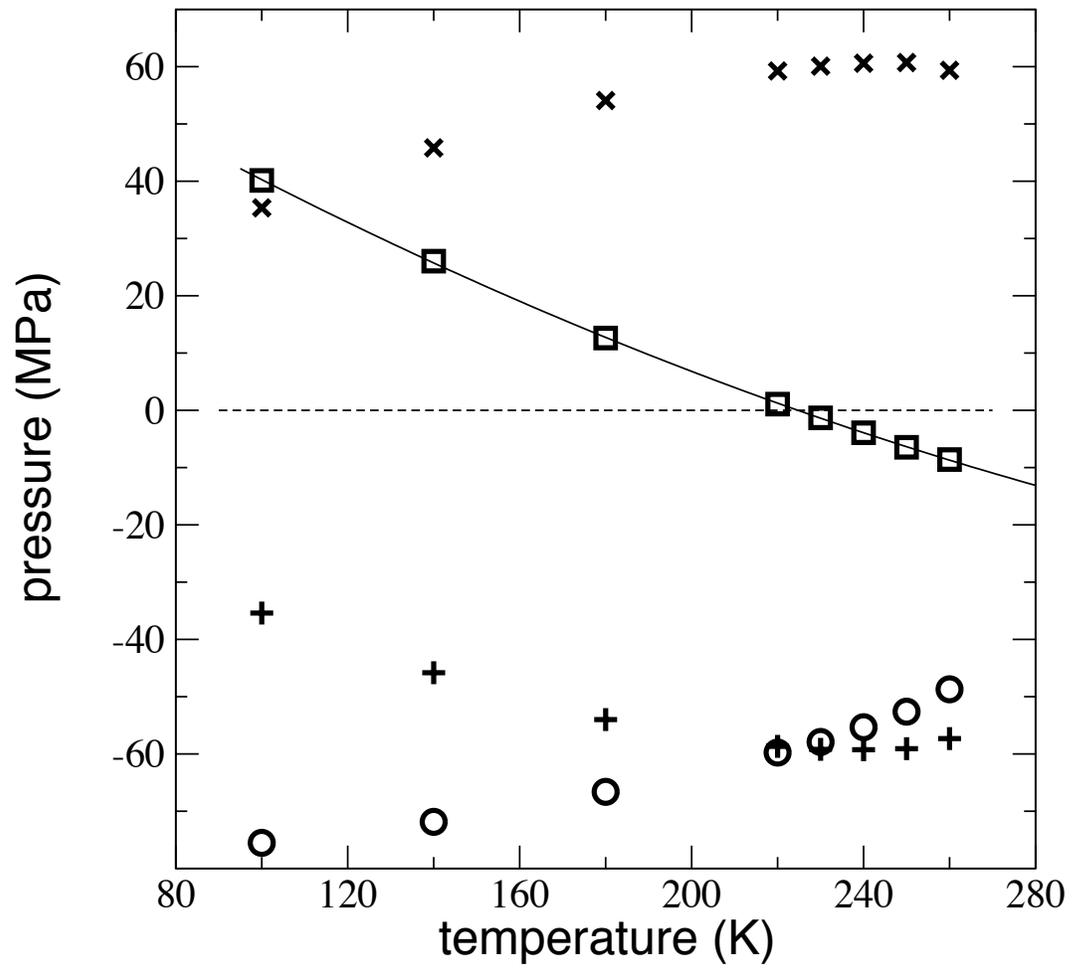

**Figure 8**

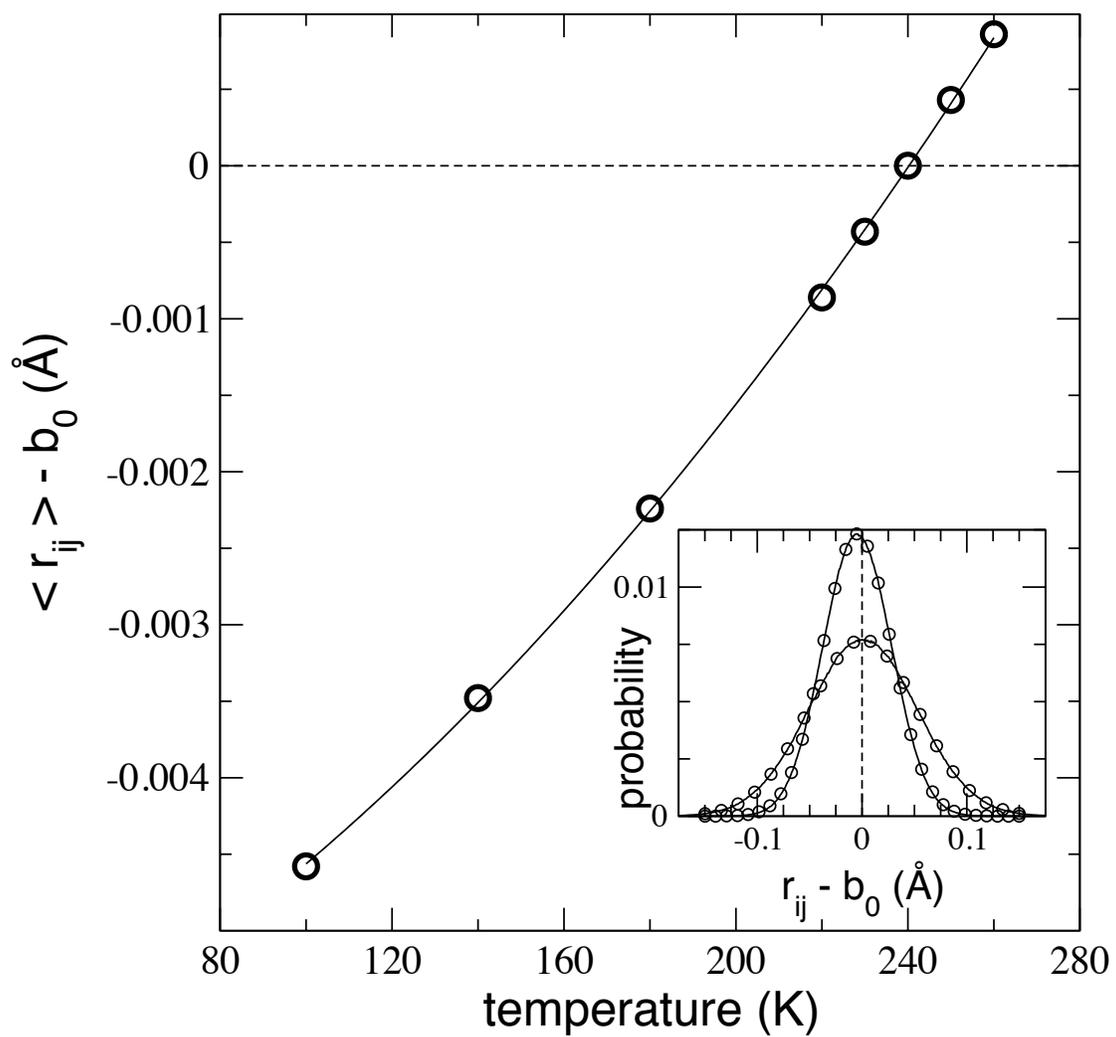

**Figure 9**

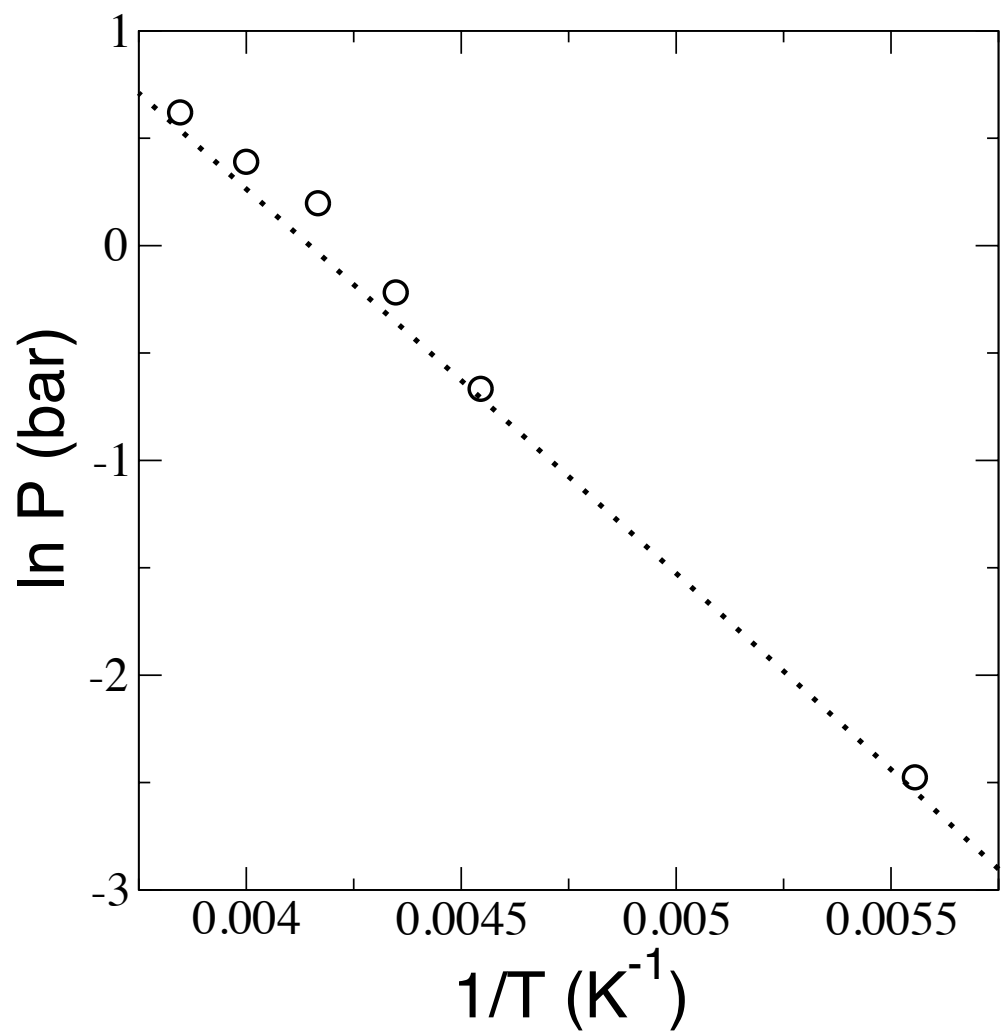